\documentclass{aa}  

\usepackage{graphicx}
\bibpunct{(}{)}{;}{a}{}{,} 
\usepackage{natbib}
\usepackage{amsmath,multirow}
\usepackage{txfonts}
\usepackage[colorlinks=true,urlcolor=blue,citecolor=blue,linkcolor=blue]{hyperref}

\usepackage{url,lineno,microtype,subcaption}
\begin{document}

  \title{Power distribution of oscillations in the atmosphere of a plage region}
    
  \subtitle{Joint observations with ALMA, IRIS and SDO}

   \author{Nancy Narang  \inst{1,2}
          	\and
       		Kalugodu Chandrashekhar \inst{1,2}
       		\and
       		Shahin Jafarzadeh \inst{1,2}
       		\and
       		Bernhard Fleck \inst{3}
       		\and
       		Miko{\l}aj Szydlarski \inst{1,2}
       		\and
       		Sven Wedemeyer \inst{1,2}
       		}

   \institute{Rosseland Centre for Solar Physics, University of Oslo, PO Box 1029, Blindern 0315 Oslo, Norway\\
              \email{nancy.narang@astro.uio.no}
         \and
             Institute of Theoretical Astrophysics, University of Oslo, PO Box 1029, Blindern 0315 Oslo, Norway
             \and
             ESA Science and Operations Department, c/o NASA Goddard Space Flight Center, Greenbelt, MD 20771, USA
             }


 
  \abstract
   {Joint observations of Atacama Large Millimeter/Submillimeter Array (ALMA) with other solar observatories can provide a wealth of opportunities to understand the coupling between different layers of the solar  atmosphere.}
   {In this article, we present a statistical analysis of power distribution of oscillations in a plage region in active region NOAA AR12651, observed jointly with ALMA, IRIS (Interface Region Imaging Spectrograph), and SDO (Solar Dynamics Observatory).}
   {We employ coordinated ALMA Band-6 (1.25\,mm) brightness temperature maps, IRIS Slit-Jaw Images in 2796\,\AA~passband, and observations in six passbands (1600\,\AA, 304\,\AA, 131\,\AA, 171\,\AA, 193\,\AA\,and 211\,\AA) of AIA (Atmospheric Imaging Assembly) onboard SDO. We perform Lomb-Scargle transforms to study the distribution of oscillation power over the observed region by means of dominant period maps and power maps. We study spatial association of oscillations through the atmosphere mapped by the different passbands, with focus on the correlation of power distribution of ALMA oscillations with others.}
   {We do not observe any significant association of ALMA oscillations with IRIS and AIA oscillations. While the global behavior of ALMA dominant oscillations shows similarity with that of transition region and coronal passbands of AIA, the ALMA dominant period maps and power maps do not show any correlation with those from the other passbands. The spatial distribution of dominant periods and power in different period intervals of ALMA oscillations is uncorrelated with any other passbands.}
   {We speculate the non-association of ALMA oscillations with those of IRIS and AIA be due to significant variations in the height of formation of the millimeter continuum observed by ALMA. Additionally, the fact that ALMA directly maps the brightness temperature, in contrast to the intensity observations by IRIS and AIA, can result in the very different intrinsic nature of the ALMA oscillations compared to the IRIS and AIA oscillations.}

   \keywords{Sun: atmosphere, Sun: faculae, plages, Sun: oscillations, Sun: radio radiation, Sun: UV radiation
               }

   \maketitle
%

\section{Introduction}

  \label{intro}
  
    The solar chromosphere is a vast reservoir of magnetohydrodynamic (MHD) wave energy with numerous complex structures showing a variety of oscillatory phenomena over a wide range of magnetic environments \citep{Narain90, Narain96, Morton12, Jess15}. The convective flows below the solar surface and dynamics of the solar photosphere and chromosphere excite and manifest themselves in the form of various magnetoacoustic oscillations in these layers \citep[e.g.][]{Brynildsen2003, Morton2013, Jafarzadeh2013, Gafeira2017, Stangalini2017}. These ubiquitous oscillations are coupled in several ways within the solar atmospheric layers and are one of the candidates for chromospheric and coronal heating \citep{Nakariakov2005, Taroyan09, Wedemeyer2009, Arregui2015, Gilchrist2021}.
  
    The chromosphere, being the bridge between the relatively cool photosphere and the intensely hot corona, is the key region to understand the heating mechanisms involved \citep{Carlsson2019}. The chromosphere is brightest in plage regions, which mostly are the chromospheric imprints of the foot-point regions of coronal loops and thus serve as efficient conduits in mass and energy cycle of the atmosphere \citep{Carlsson2015}. A large variety of oscillatory phenomena has been observed in plage regions, exhibiting a range of periods, extending from tens of seconds to up to few minutes (\citealt{Depontieu2007, Morton2014A, Morton2014B, Gafeira2017, Jafarzadeh2017, Kayshap2020}, also see the review by \citealt{Jess15}).

    \begin{figure*}[htbp]
    	\centering
    	
    	\includegraphics[width=0.95\textwidth]{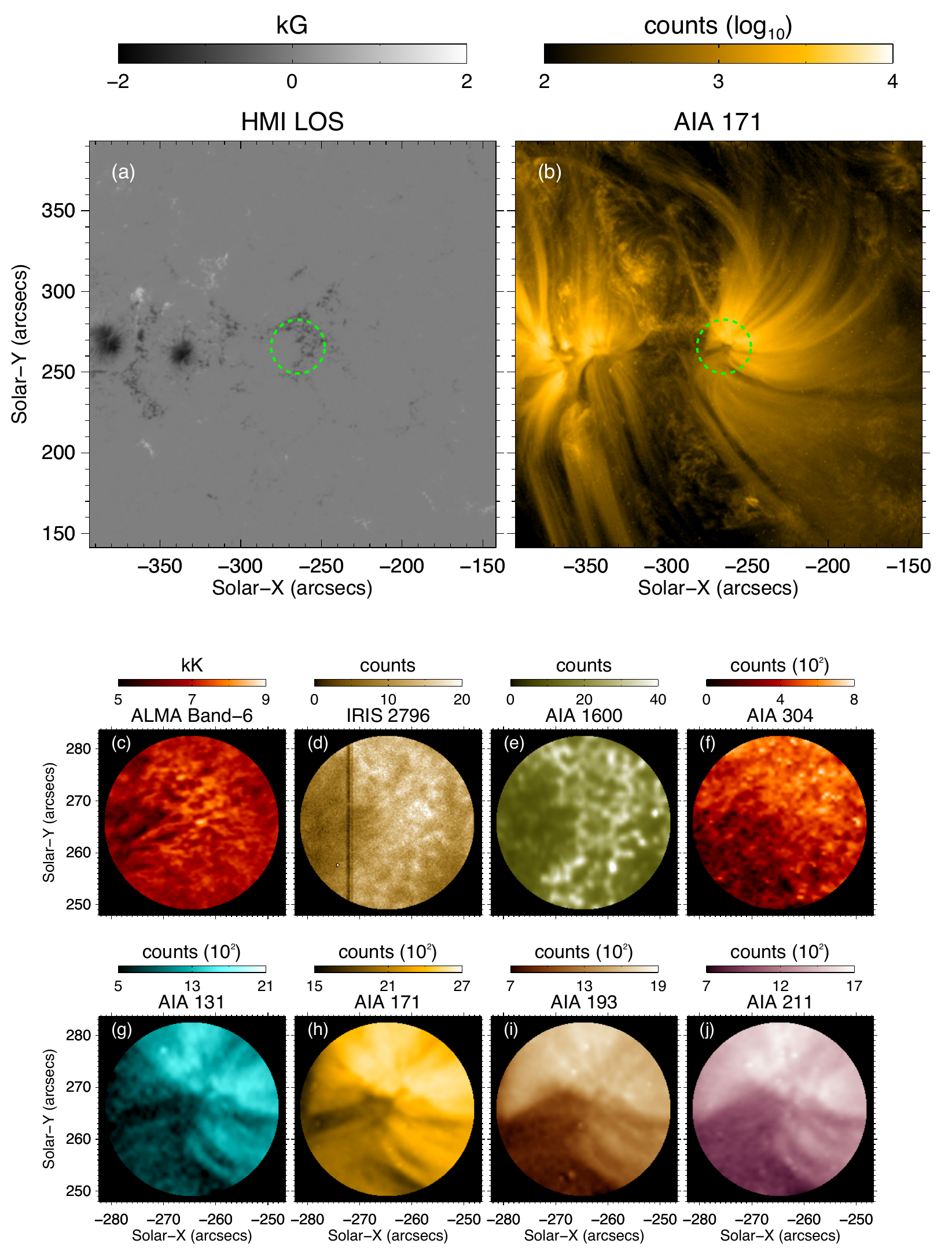}
    	
    	\caption{Context images showing HMI line-of-sight (LOS) magnetogram in panel (a) and AIA\,171\,\AA~image in panel (b) at the start time of the ALMA observations. The FoV studied here is marked by the \textit{green} circle in panels (a) and (b). Panels (c) to (j) show the representative images of the studied FoV (from ALMA Band-6, IRIS SJI\,2796\,\AA~, and different AIA channels as indicated on top of the panels) at the start time of the observations.}
    	
    	\label{fig1}
    \end{figure*}

    In the present article, we describe the analysis of distribution of oscillations over a range of periods in an active region (NOAA AR12651) plage observed jointly with the Atacama Large Millimeter/Submillimeter Array (ALMA, \citealt{Wooten2009}), IRIS (Interface Region Imaging Spectrograph, \citealt{Depontieu14}, and AIA (Atmospheric Imaging Assembly, \citealt{lemen12}) onboard the Solar Dynamics Observatory (SDO). Such coordinated solar observations of ALMA with IRIS and SDO provide a unique opportunity to study the solar atmosphere at millimeter wavelengths in conjunction with the ultraviolet part of the solar spectrum. ALMA observes the continuum emission in the millimeter wavelength range, and provides a direct measurement of gas temperature that is equivalent to the brightness temperature recorded.

    Solar observations with ALMA have, so far, mostly been taken in Band-3 (centered at around 100\,GHz\,; 3\,mm) and Band-6 (centered at around 239\,GHz\,; 1.25\,mm), which are predicted to sample the mid-to-high chromosphere \citep{Wedemeyer2016, Bastian2017, Nindos2018, Jafarzadeh2019, Loukitcheva2019, Martinez2020}. Recent studies have also indicated that the ALMA observations may have some contributions from transition region and lower coronal heights \citep{Wedemeyer2020, Chintzoglou2021A, Chintzoglou2021B}. For more details about the solar observations with ALMA and recent results, we refer the reader to the reviews by \citet{Wedemeyer2016} and \citet{Loukitcheva2019}.

    In this study, we explore the presence of possible associations of ALMA observations with chromospheric and lower coronal observations from IRIS and AIA from the point of view of oscillations. The plage region studied is shown in the Fig.~\ref{fig1}, within the green circle, on a larger field-of-view (FoV) of a representative HMI (Helioseismic and Magnetic Imager, \citealp{Scherrer12}) line-of-sight magnetogram and AIA\,171\,\AA~image in the top panels. The first frames of the studied FoV of the different passbands are shown in the panels in the bottom. The article is structured as follows: in Sect.~\ref{obs} we provide details of the joint observations, data-processing and co-alignment techniques applied. In Sect.~\ref{data} we explain the data-analysis technique used (Lomb-Scargle Transform). We present our results in~\ref{res}, discussion in Sect.~\ref{disc} and conclusions in Sect.~\ref{conc}.

\section{Observations}
\label{obs}

\begin{figure}[htbp]
    	\centering
    	
    	\includegraphics[width=0.47\textwidth]{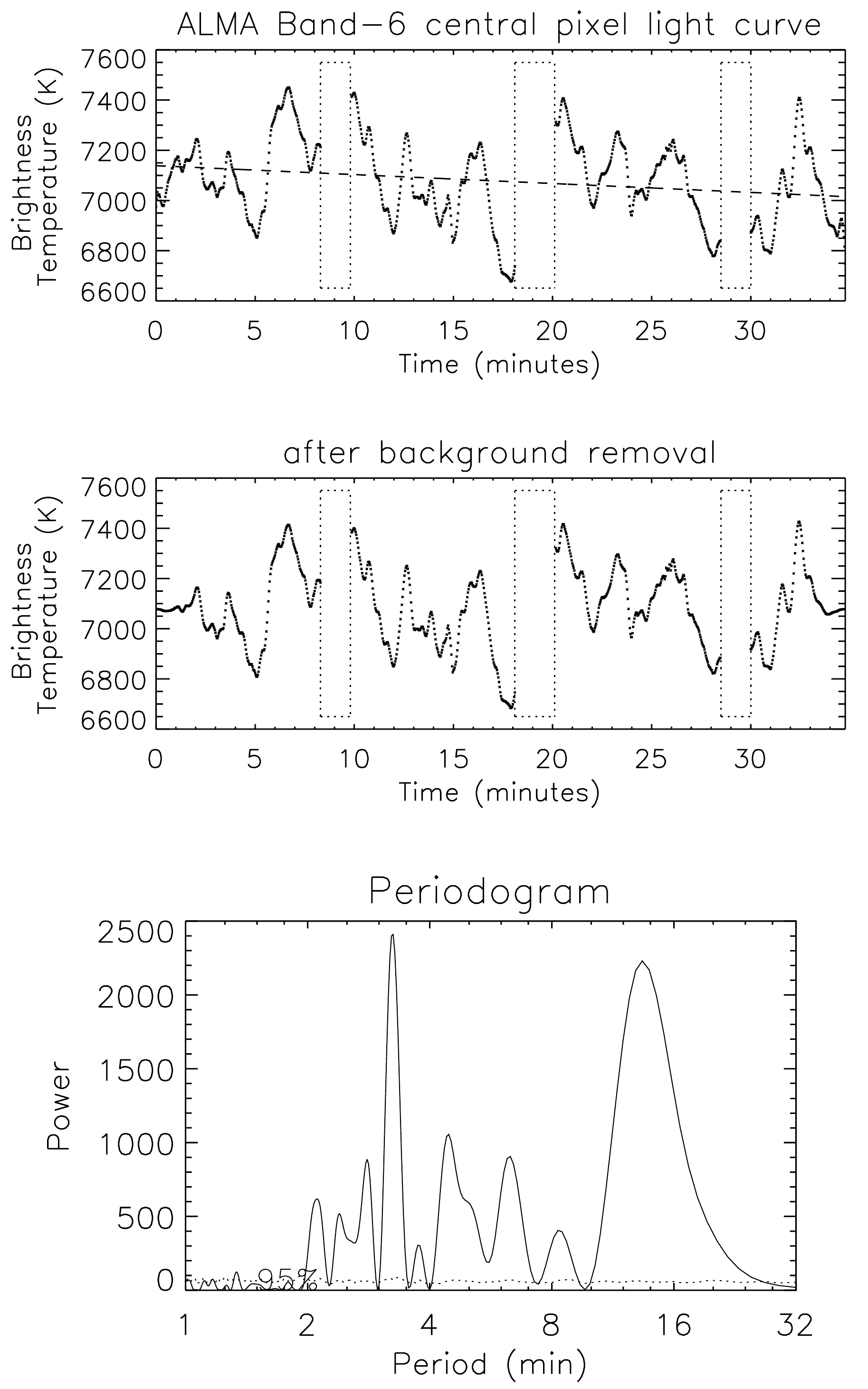}
    	
    	\caption{Lomb-Scargle analysis for the ALMA Band-6 brightness temperature variation with time at the pixel location at the center of the FoV. Top panel: light curve of the ALMA brightness temperature (black symbols) with the background trend (dashed line), and data-gaps marked within dotted boxes. Middle-panel: apodized and de-trended light-curve. Bottom panel: periodogram/power spectrum (solid line) with 95\% confidence level (dotted line). More details about the different panels are mentioned in the text (Sect.~\ref{data}).}
    	
    	\label{fig2}
    \end{figure}

    This work primarily utilizes ALMA cycle 4 solar observations of a plage region in NOAA AR12651 (project 2016.1.00050.S) in Band-6 (centered at around 239\,GHz\,; 1.25\,mm). The target, centered at (x,y)=(-265",265"), was sampled between 15:59-16:34 UT on 2017 April 22 with a 2-sec cadence in 4 blocks of approximately 8--9\,min duration each, with gaps of 1.5--2\,min in between for calibrations. See the first panel of Fig.~\ref{fig2} where the time-sequence of ALMA observations (brightness temperature light-curve) at one pixel location is shown. The light-curve shows the presence of the three episodes of data-gaps, that are enclosed within the dotted rectangles.

    The ALMA dataset was calibrated using the Solar ALMA Pipeline (SoAP, Szydlarski\,et\,al. \textit{in prep.}; also see \citealt{Wedemeyer2020} for more details), which has been developed based on the Common Astronomy Software Applications (CASA) package \citep{McMullin2007}. The calibrations include the CLEAN algorithm as implemented by \citet{Rau2011}, self-calibration for a time window of 14\,sec, primary beam correction, and combination of the interferometric data with the total power maps ("feathering" procedure). The above mentioned calibration steps finally provide the ALMA brightness temperature maps. For further details about SoAP, also see \citet{Jafarzadeh2019, Jafarzadeh2021}.

    The ALMA observations were supported by co-observations of the same region by IRIS. For further details about the joint ALMA and IRIS observations, we refer the reader to \citet{Santos2020} and \citet{Chintzoglou2021A, Chintzoglou2021B} in which the same coordinated ALMA and IRIS observations are analyzed. This coordinated dataset is one of the first campaigns where joint observations between ALMA and IRIS were conducted, and good alignment between the ALMA and IRIS datasets has been achieved \citep{Henriques2021}. Considering the continuous observations by AIA/SDO in a wide range of wavelengths along with the joint ALMA and IRIS observations, in this work we have been able to sample the oscillations in the plage region at multiple atmospheric heights from the solar chromosphere to the corona.

\begin{figure*}[htbp]
	\centering
	
	\includegraphics[width=0.95\textwidth]{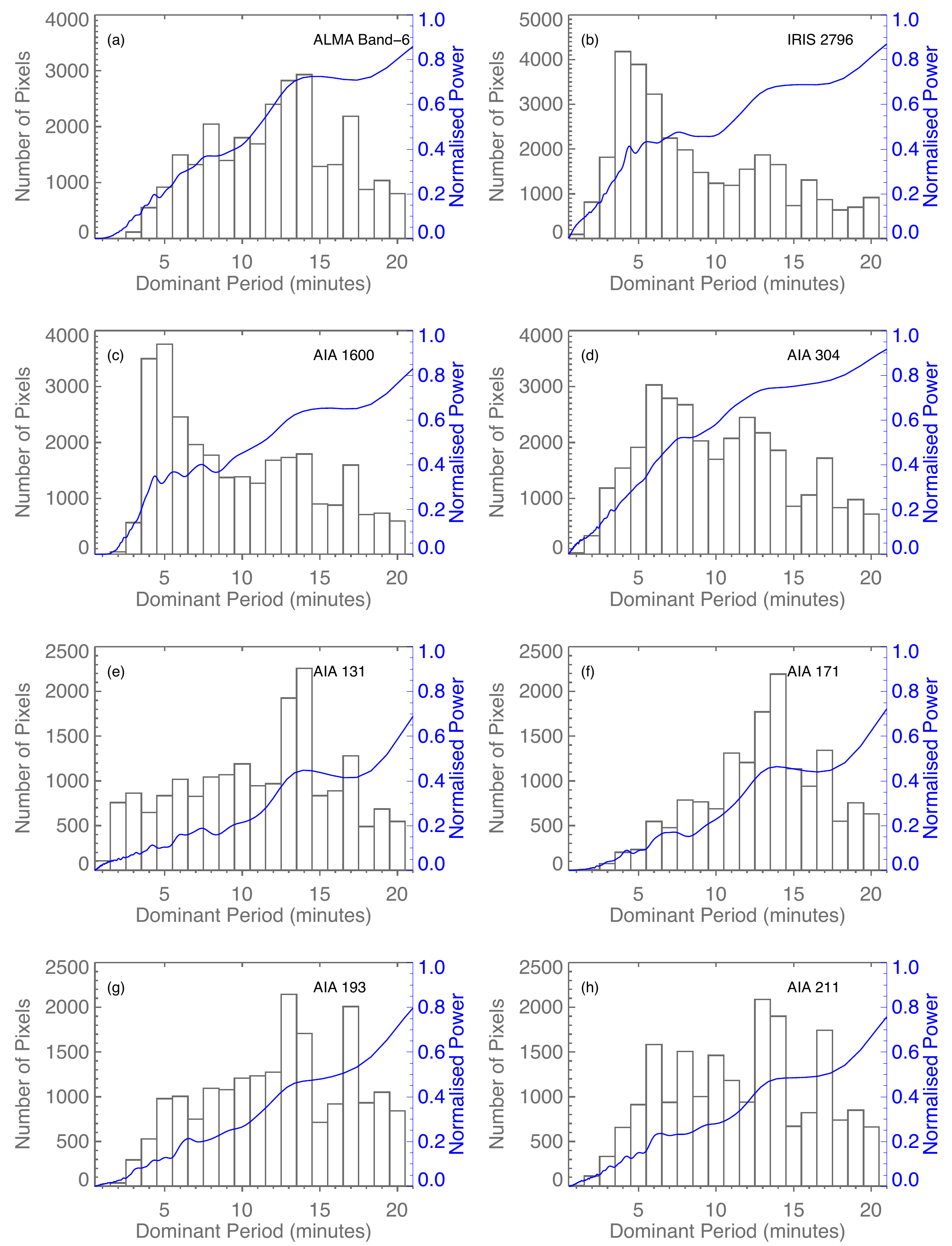}
	
	\caption{Histograms of dominant periods (grey) detected over the FoV in the different passbands. Overplotted in blue are the average power spectra for the respective passband.}
	
	\label{fig3}
\end{figure*}

    In the present work we have used AIA observations in 1600\,\AA, 304\,\AA, 131\,\AA, 171\,\AA, 193\,\AA and 211\,\AA~passbands, and IRIS\,2796\,\AA~slit-jaw images (SJIs) with the ALMA Band-6 observations (see Fig.~\ref{fig1}). All the images from IRIS and AIA have been aligned (spatially and temporally) with those from ALMA images/brightness temperature maps. For this, the IRIS and AIA images were re-scaled to the same sampling resolution of ALMA images, i.e., 0.14 arcsec, prior to spatial alignments to the same field of view. The images from the IRIS and AIA were, however, analyzed with their original spatial resolution (\textit{i.e.}, they were not convolved with the PSF of ALMA). To match the higher cadence of the ALMA images, the images in the utilized IRIS and AIA time series were repeated in time, when necessary.

\section{Data analysis}
\label{data}

    We employ Lomb-Scargle (LS) transforms \citep{Scargle82, Press89} to detect and characterize the oscillations present in ALMA Band-6 brightness-temperature maps of the plage region. This method, in particular, is suitable for cases where the observation times are unevenly spaced, like in the present ALMA observations (as mentioned in Sect.~\ref{obs}, also see Fig.~\ref{fig2}). We use the standard \textit{IDL} routine \textit{LNP\_TEST} to perform LS transforms with the statistical significance test of the detected periodicities against the hypothesis of presence of random/white-noise in the input signal.

    LS transform is performed at each pixel location in the ALMA FoV to extract the information about the distribution of power in different periodicities over the observed plage region. Figure~\ref{fig2} shows a representative example of the results from the LS analysis corresponding to the pixel location at the center of the FoV. The top panel in Fig.~\ref{fig2} shows the variation of the ALMA brightness temperature (light-curve) with time. The instances of data-gaps along the light-curve are enclosed within the dotted boxes in the top and middle panels.

     The middle-panel shows the light-curve after removal of the linear background trend (dashed line in the first panel) and apodization, that is further used to obtain the periodogram/power spectrum shown in the bottom panel. The trend subtraction helps to remove the periodicities which can arise due to non-stationary data. The data are then apodized (using a Tukey window) to remove the edge-effects. To increase the frequency/period resolution, the signal is padded with zeros (at one end) prior to the LS transform. The padding is performed such that it increases the length of the signal by a factor of 5. The bottom panel displays the periodogram/power spectrum (solid line) with 95\% confidence level (dotted line). We further use the periodograms obtained at every location in the FoV to study the spatial distribution of ALMA oscillations and its association with those from IRIS and AIA observations.

     The periodogram/power spectrum obtained in the particular example in Fig.~\ref{fig2} has a prime dominant power peak at a period of 3.3\,min. The second strong (dominant) power peak, at around 13.7\,min, is only slightly lower than that at 3.3\,min. The power spectrum also constitutes multiple weaker, yet significant, peaks. It is thus worth noting that the dominant period estimation may be biased when more than one strong peak occurs in a power spectrum. Thus, they should be interpreted with great caution (i.e., a dominant period does not necessarily represent the only dominant oscillation in a pixel, but the largest, or one of the largest, oscillation power found in that pixel). This should not, however, affect the overall statistical associations which we study in the following sections. This is elaborated further in Sect.~\ref{gb}.

\begin{figure}[htbp]
	\centering
	\includegraphics[width=0.47\textwidth]{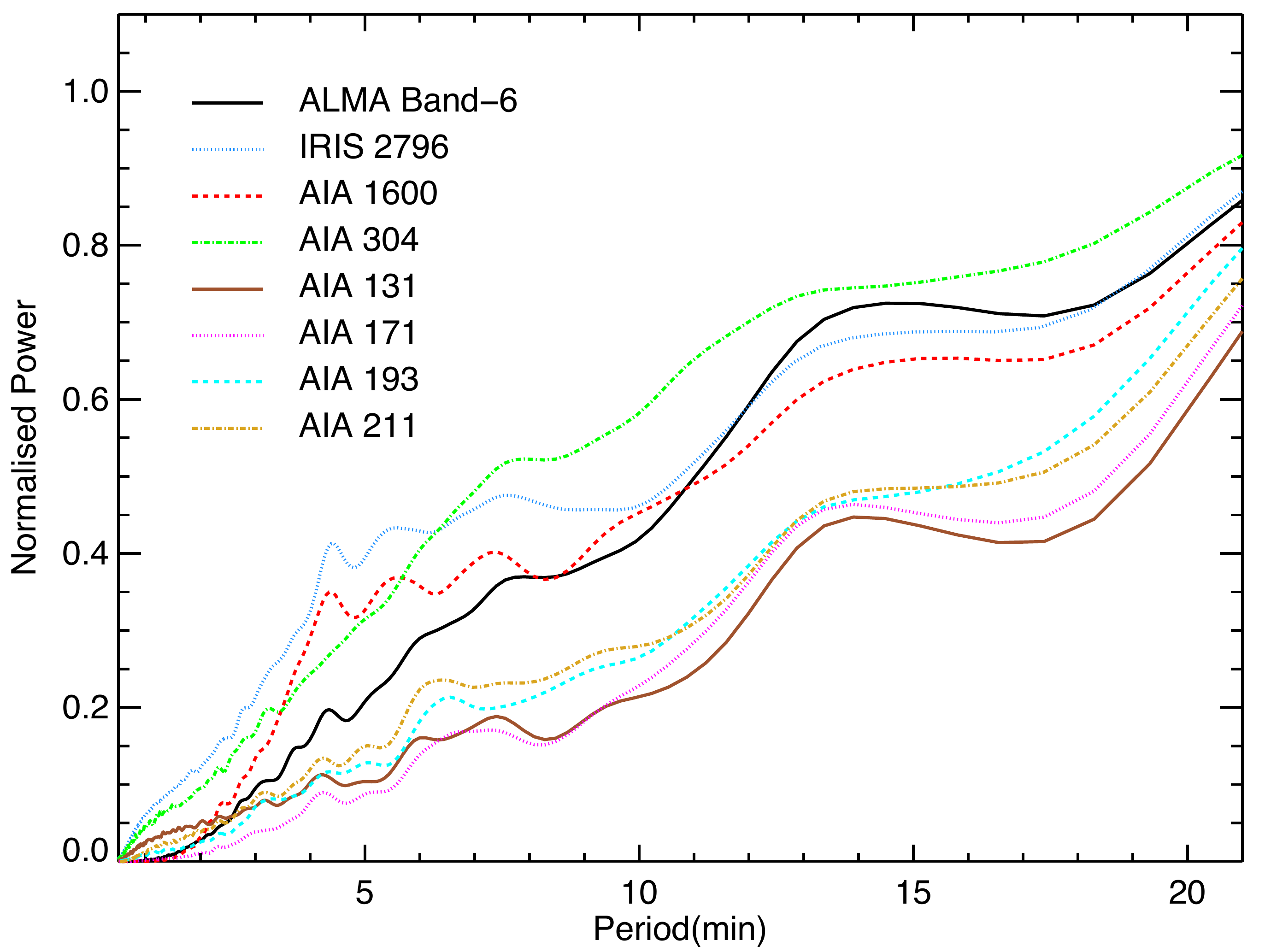}
	
	\caption{Average power spectra of the eight passbands shown in Figure~\ref{fig3}.}
	
	\label{fig4}
\end{figure}

\section{Results}
\label{res}

\subsection{Distribution of dominant period}
\label{ddp}

 \subsubsection{Global behavior}
 \label{gb}
 
    The dominant period (period at peak power in the periodogram) is determined for every pixel location in all the passbands. The global behavior of preferential dominant period in the different passbands is studied by means of histograms and average power spectra as shown in Fig.~\ref{fig3}. The different panels in Fig.~\ref{fig3} show the histograms (grey color) of the dominant periods for the different passbands. Overplotted in blue are the respective average power spectra (averaged over all power spectra determined at all individual pixels). As mentioned in Sect.~\ref{data} the dominant periods estimation could be biased at pixels where more than one dominant peak is observed in their power spectra. However, from the statistical point of view, this should likely not affect the overall global trends. This is justified in Fig.~\ref{fig3}, where the variations (appearances of peaks) in the histograms and the corresponding average spectra are in good agreement, demonstrating that the dominant periods estimated are well representing the average state of the oscillations. It should be noted that the average power spectra shows the expected rising trend with increasing period (also see Fig.~\ref{fig4}). Though we have obtained the dominant period values up to $\sim$35\,min, we restrict the analysis up to 20\,min. This is due to the fact that the longer dominant periods detected can be related to the long term evolution of the various features in the atmosphere and not to oscillations.

    The ALMA histogram in panel (a) of Fig.~\ref{fig3} shows that most of the locations in the ALMA FoV have the dominant period of oscillation in 12--14\,min period range. In contrast, the histograms of the IRIS 2796\,\AA~in panel (b) and AIA 1600\,\AA~in panel (c) (sampling heights approximately corresponding to the middle chromosphere and low chromosphere, respectively) show the prominent dominant oscillations to be in 4--6\,min period range. This contrast is also evident from Fig.~\ref{fig4}, where the power spectra for all the passbands are shown together for a better comparison. It is important to note here that we are unable to perform any comparisons of ALMA oscillations with those present in the upper chromospheric layers due to the following limitation of the observations. The upper chromosphere/transition region can typically be observed by IRIS SJIs 1330\,\AA~and 1400\,\AA~passbands. Though the dataset used here contain the SJ observations in these passbands, the signal in these passbands is poor in this observation set to be considered for any time-frequency transform such as Lomb-Scargle. These SJ observations are underexposed with an average count of two and four in 1300\,\AA~and 1400\,\AA~passband, respectively.

  \begin{figure*}[htbp]
  	\centering
  	
  	\includegraphics[width=0.94\textwidth]{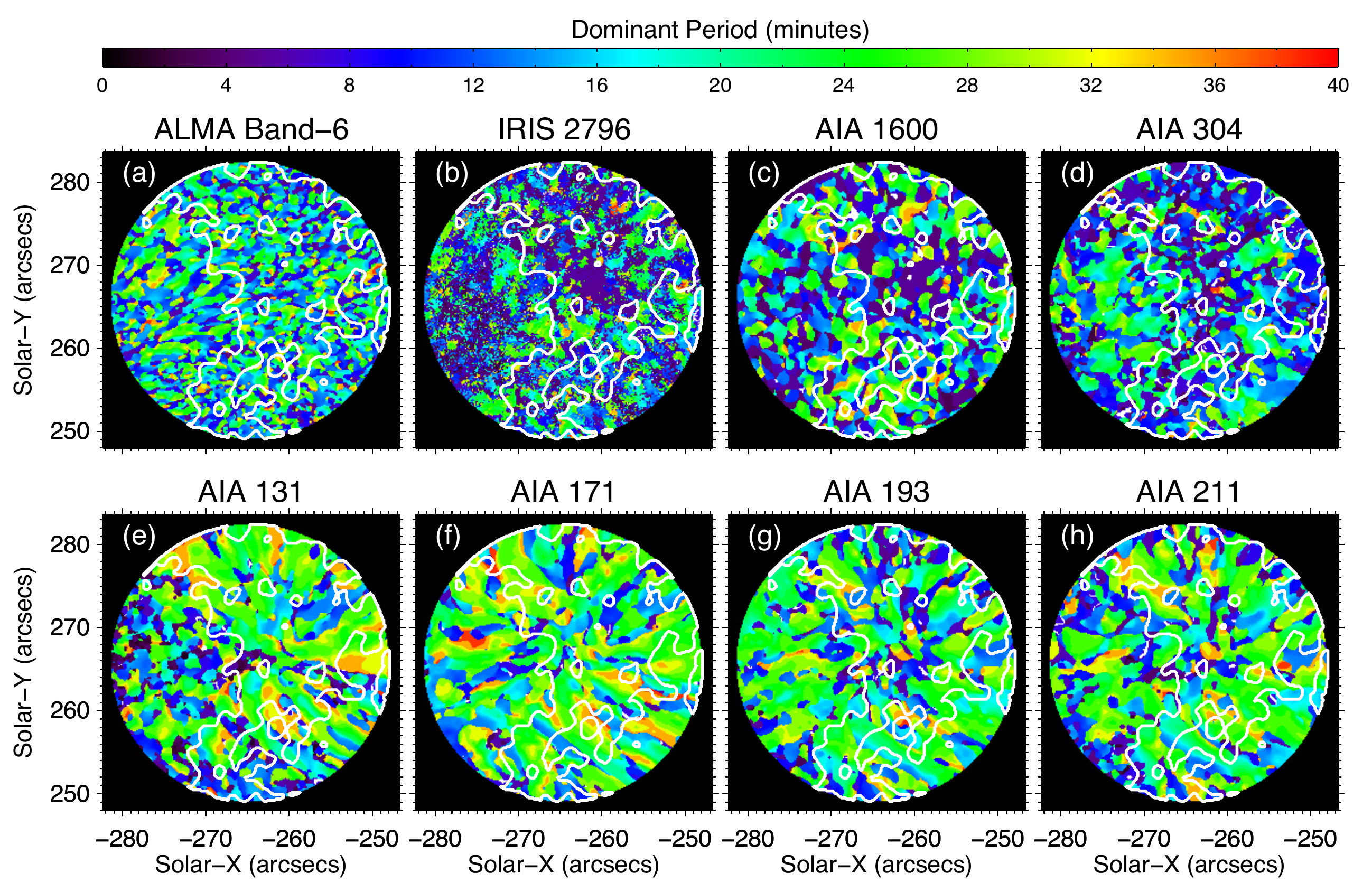}
  	
  	\caption{Dominant period maps of the different passbands studied.}
  	
  	\label{fig5}
  \end{figure*}

  \begin{figure*}[htbp]
  	\centering
  	
  	\includegraphics[width=0.94\textwidth]{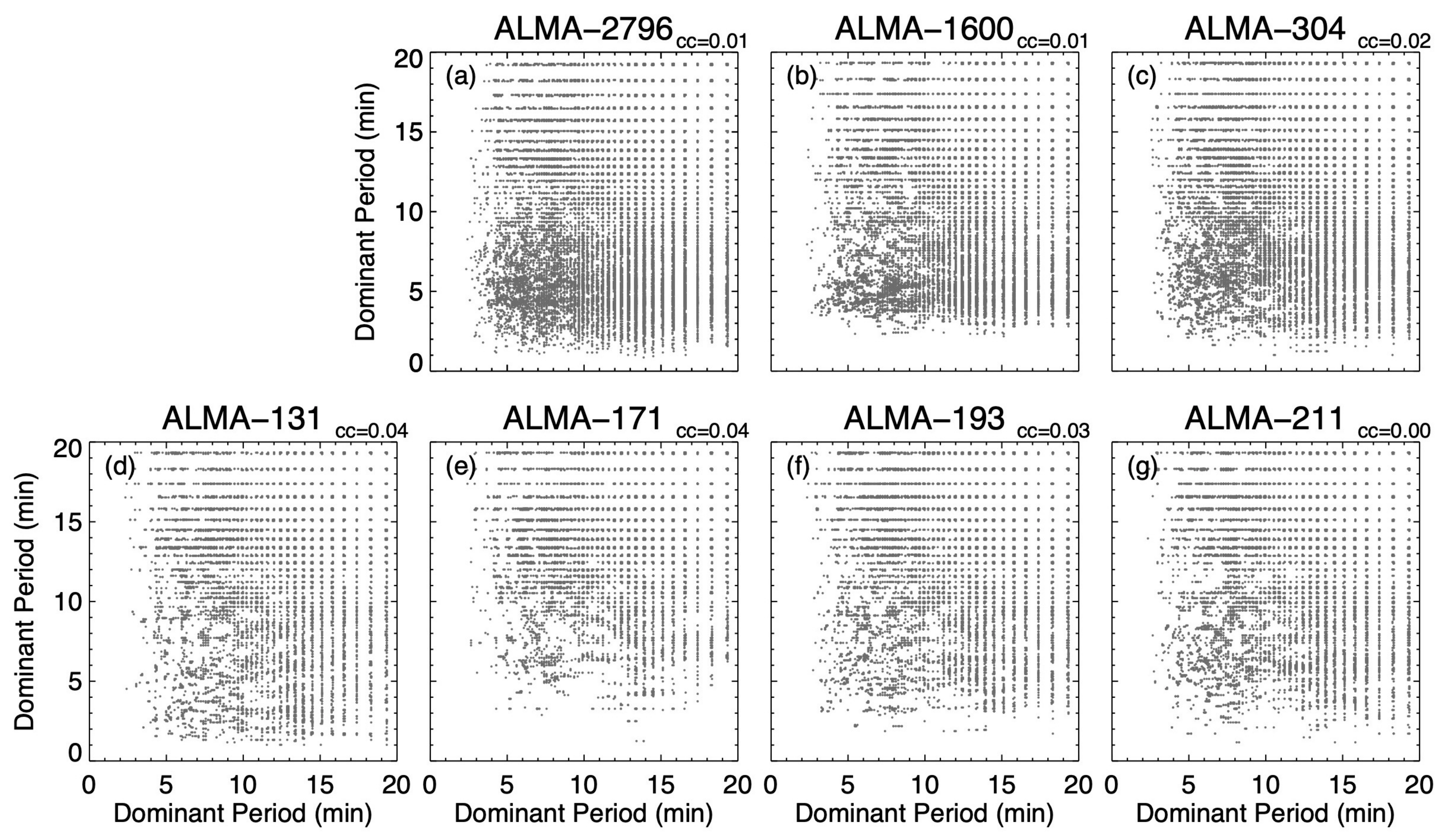}
  	
  	\caption{Scatter plots, for the full FoV, of ALMA Band-6 dominant period map with that from IRIS\,2796\,\AA~in panel (a), and different AIA channels in panels (b) to (g). cc represents the value of the cross-correlation coefficient.}
  	
  	\label{fig6}
  \end{figure*}

  \begin{figure*}[htbp]
  	\centering
  	
  	\includegraphics[width=0.94\textwidth]{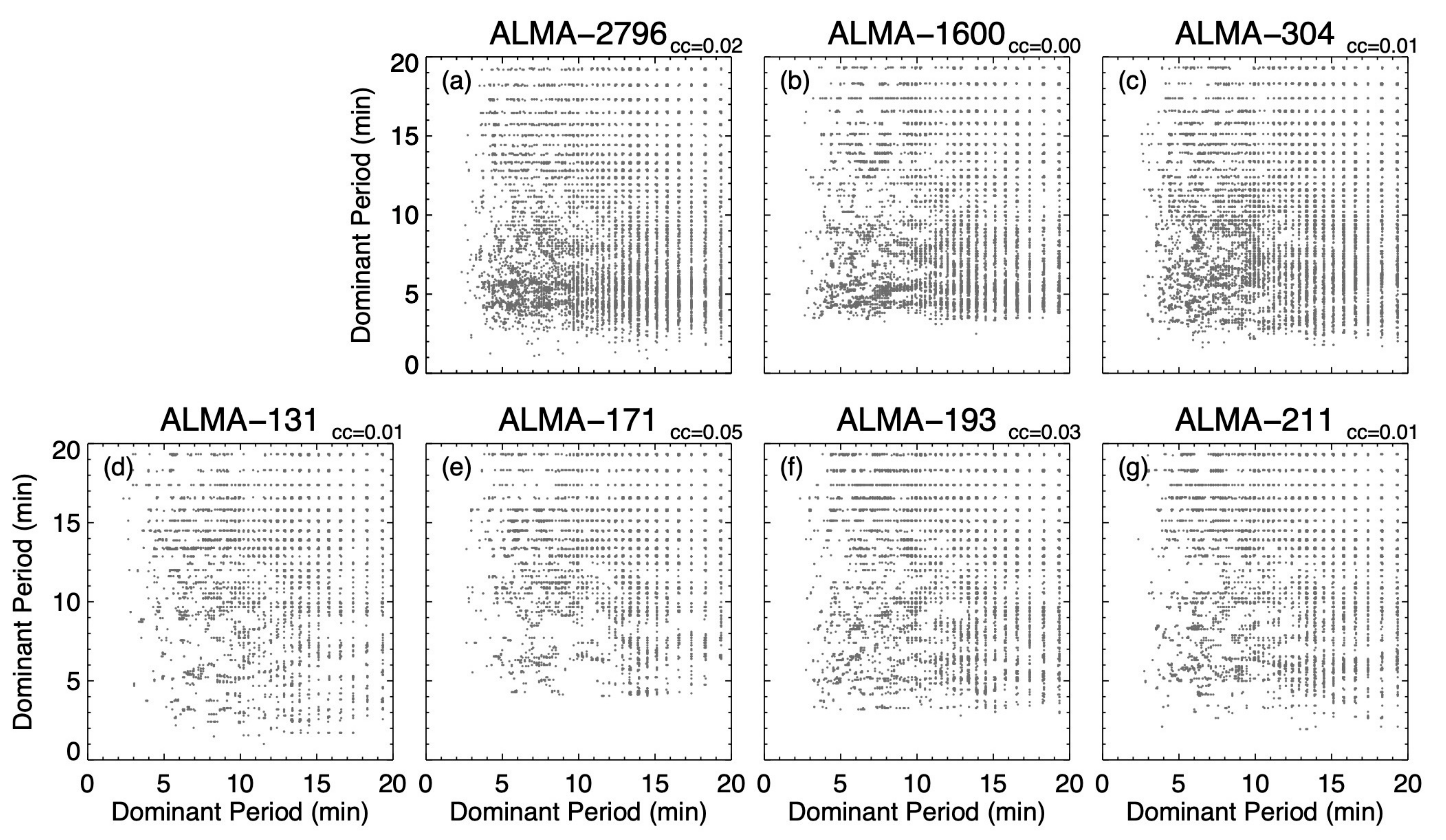}
  	
  	\caption{Same as figure~\ref{fig6} but only for the bright plage region of the FoV.}
  	
  	\label{fig7}
  \end{figure*}

  \begin{figure*}[htbp]
  	\centering
  	
  	\includegraphics[width=0.94\textwidth]{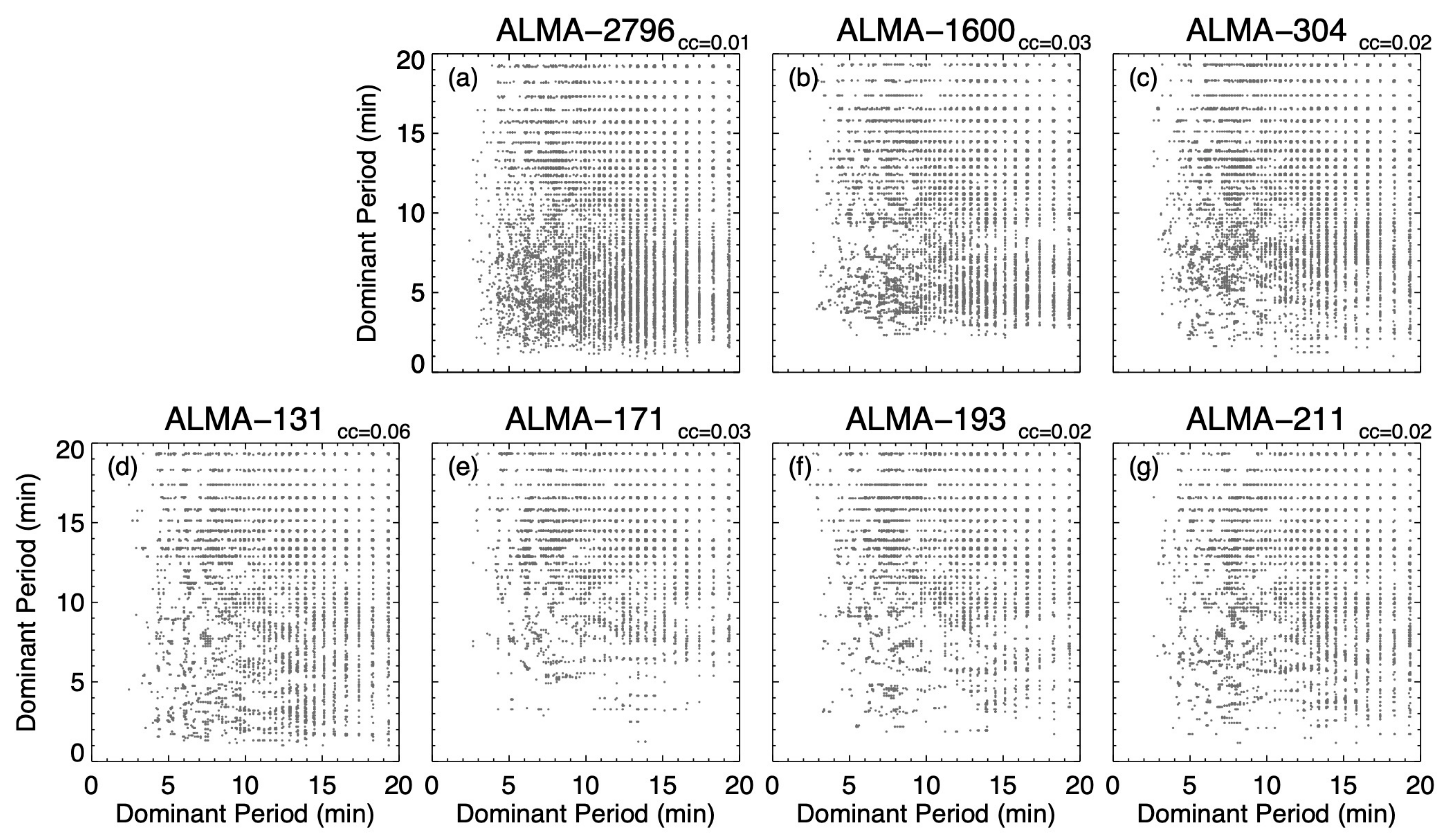}
  	
  	\caption{Same as figure~\ref{fig6} but only for the peripheral region of the FoV.}
  	
  	\label{fig8}
  \end{figure*}

    The oscillations with $\sim$3\,min period in the chromosphere are well understood and are explained as the basic cut-off frequency resonance of the chromosphere \citep{Deubner90, Fleck91, Carlsson94, Rutten95}. As mentioned in Sect.~\ref{intro} (and citations therein), various numerical models and simulations have predicted ALMA to probe plasma conditions in the middle to high chromosphere. The very different global behavior of ALMA oscillations observed here, indicate the need for more detailed investigations to determine the formation height of the radiation observed by ALMA. The absence of dominant 3--5\,min oscillations in ALMA observations has been discussed in \citet{Jafarzadeh2021} (a more detailed study will be presented in Jafarzadeh\,et\,al. \textit{in prep.}).

    The transition region passband of AIA 304\,\AA ~can have contributions from the higher chromosphere up to the lower corona. The histogram for AIA 304\,\AA ~in panel (d) of Fig.~\ref{fig3} shows two peaks; while the one at 6--8\,min is prominent, the other at 12--14\,min is also comparable. The peak around 12--14\,min is also present in the histograms of all the coronal passbands of AIA in panels (e) to (h). As the ALMA histogram in panel (a) also shows a prominent dominant period in the 12--14\,min range, it can be conjectured that the global behavior of ALMA oscillations may have an association with the transition region and lower coronal oscillations, rather than with the chromosphere. The presence of possible spatial correlations, if any, are further investigated in more detail in the following sections.

 \subsubsection{Spatial correlation}
 \label{sc}
 
    We further study the association of dominant periods in ALMA oscillations with those of IRIS and AIA by means of dominant period maps. A dominant period map shows the spatial distribution of the dominant period over the FoV. The dominant periods (period at peak power in the periodogram) are determined for every pixel location for all passbands and are shown in Fig.~\ref{fig5} for the entire FoV of the different passbands. By means of scatter plots we study the correlation of the ALMA dominant period map with the other passbands. Figure~\ref{fig6} shows the different scatter plots with the value of cross-correlation coefficients (cc) between the respective full FoV dominant period maps. We do not observe any statistical trends or patterns in the scatter plots.

    We further partitioned the FoV into two sub-regions, the bright plage region and the peripheral region, based on the intensity threshold from AIA 1600\,\AA~observations (see Appendix~\ref{apndxA} for details). The \textit{white} contours in Fig.~\ref{fig5} enclose the bright plage region, and the surrounding area (outside the contours) is termed as the peripheral region. The association among the dominant period is then further studied, in the similar manner as mentioned above, separately in the bright plage region and the peripheral region. The dominant period histograms and average LS power spectra obtained separately for bright plage and peripheral region do not show different trends for respective passbands, which are also similar to the corresponding ones from the full FoV shown in Figs.~\ref{fig3} and~\ref{fig4}. Figures~\ref{fig7} and \ref{fig8} show the scatter plots between the ALMA dominant period map with the other passbands, for the bright plage region and the peripheral region, respectively. The scatter plots in Figs.~\ref{fig6},~\ref{fig7}, and~\ref{fig8} show the absence of any spatial correlation of the dominant oscillations in ALMA observations with those in other passbands. The very low values of cc (less than 6\%), for all cases, indicate that the distribution of dominant periods of the ALMA oscillations does not have a direct association with those of IRIS and AIA (in these particular observations).

    The analysis above, involving the dominant period maps, holds a possibility of hiding some correlations, particularly for the short period oscillations. As mentioned earlier, the longer dominant periods (larger than 20\,min) in the dominant period maps could be arising due to the long-term evolution of the various features, that can be unrelated to each other in different atmospheric heights. This can result in low values of the cross-correlation. To avoid the effects of the long-term evolution, we further study the association of ALMA oscillation power with other passbands, separately in various period bins, by means of power maps. The following analysis in Sect.~\ref{dop} also enables us to study the correlation of the oscillations in the respective period intervals, in contrast to this section, where we only studied the dominant oscillation periods.

\subsection{Distribution of oscillation power}
\label{dop}

\begin{figure*}[htbp]
	\centering
	
	\includegraphics[width=0.90\textwidth]{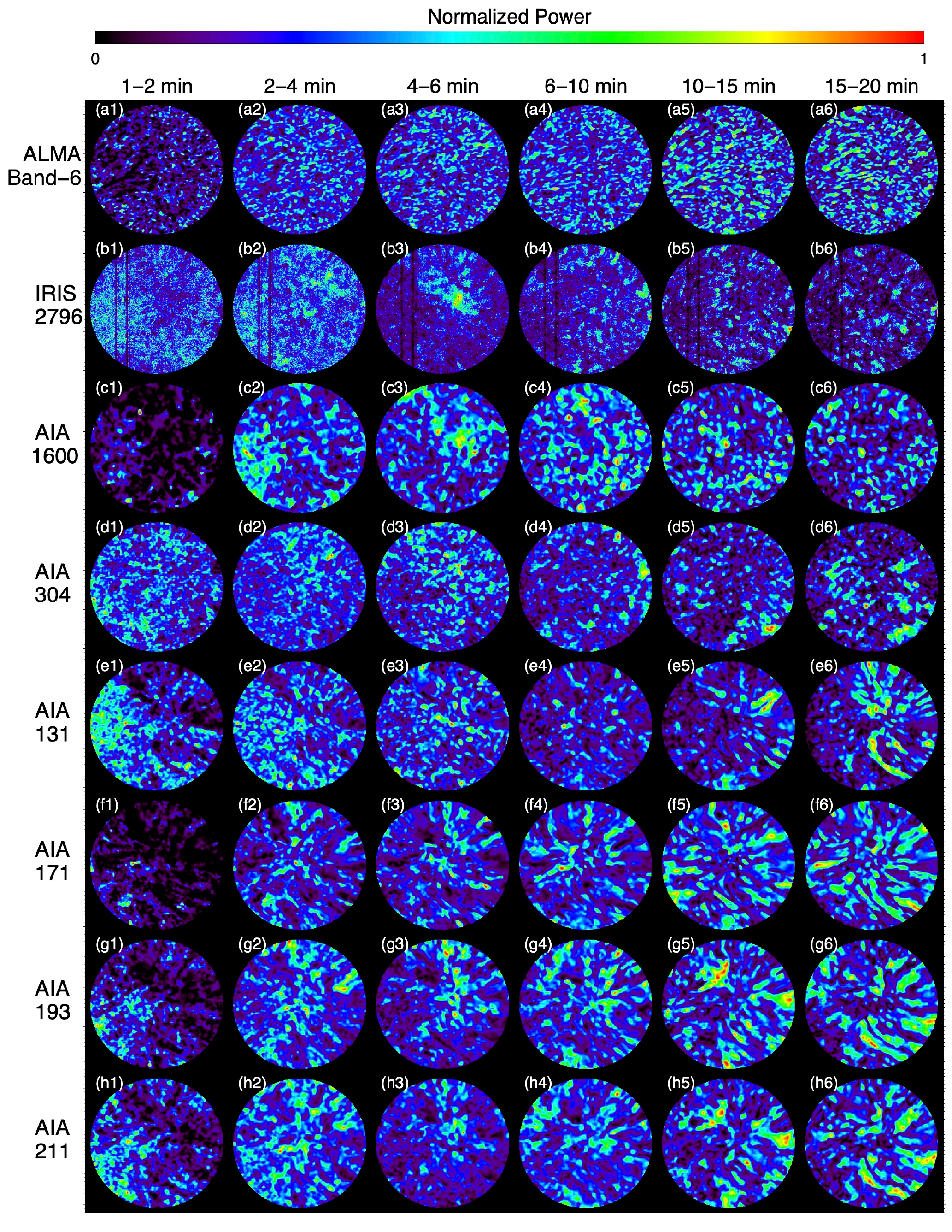}
	
	\caption{Power maps of different passbands in the six period intervals.}
	
	\label{fig9}
\end{figure*}

    To further explore the inter-relation of ALMA oscillations with the other chromospheric, transition region and lower coronal observations, we study the association of the distribution of ALMA oscillation power with that of the IRIS and AIA oscillations. The power maps show the value of the power averaged over the desired period interval in the periodogram at each pixel location over the FoV. We obtain power maps in six period bins (1--2\,min, 2--4\,min, 4--6\,min, 6--10\,min, 10--15\,min, and 15--20\,min) for all datasets. Figure~\ref{fig9} shows the power maps for ALMA, IRIS 2796\,\AA, and different passbands of AIA in the six period bins. Since we are interested in the comparison of power distributions within respective period bins of different datasets, every power map is normalized to its maximum value.

\begin{figure*}[htbp]
	\centering
	
	\includegraphics[width=0.95\textwidth]{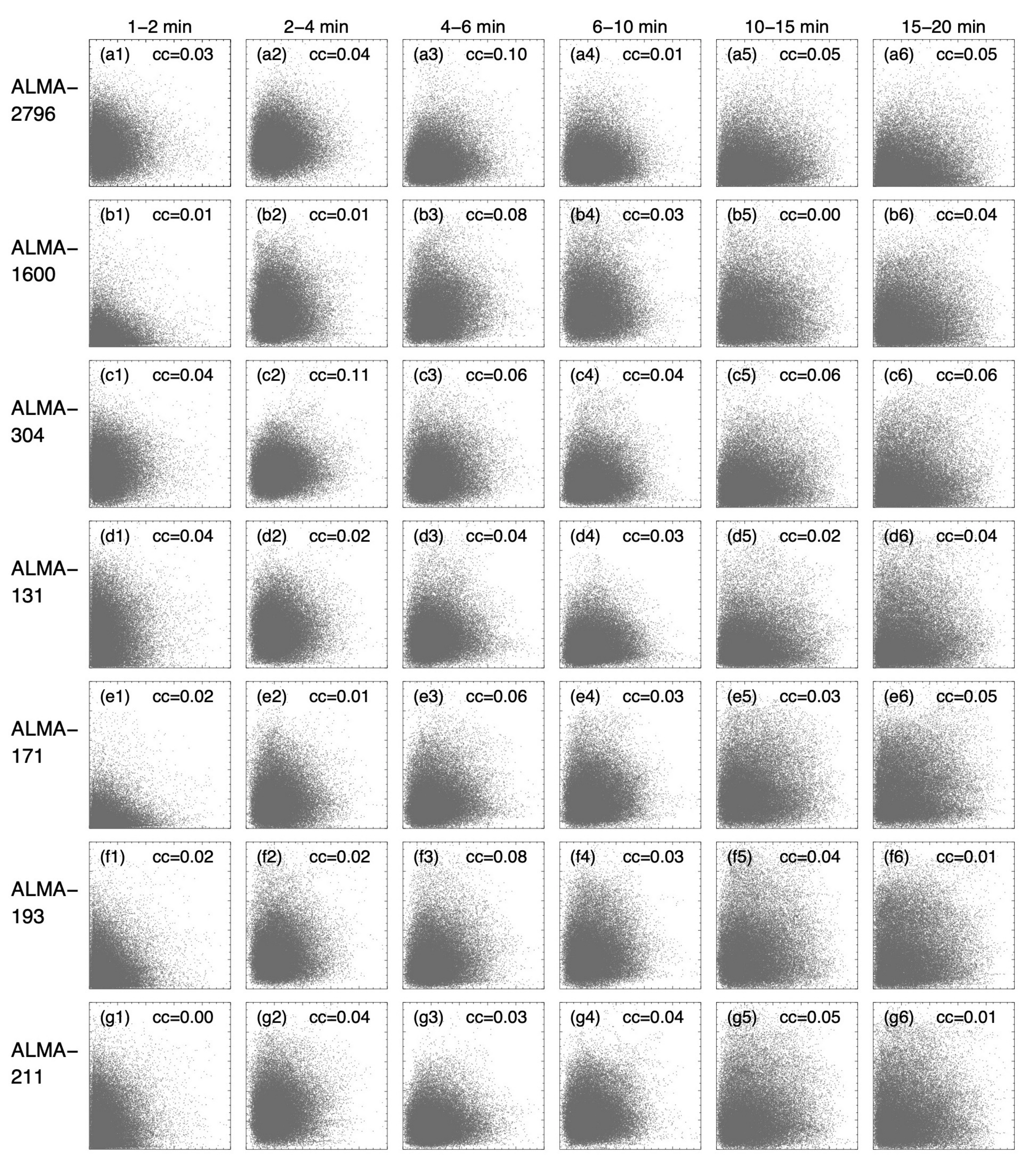}
	
	\caption{Scatter plots, for the full FOV, of ALMA Band-6 power maps in the six period intervals, with that from IRIS\,2796\,\AA~ in panels (a(1) to a(4), and different AIA channels consecutively in panels from b(1) to g(4). cc represents the value of the cross-correlation coefficient.}
	
	\label{fig10}
\end{figure*}

    The power maps of ALMA do not show any close association with the power maps of IRIS 2796\,\AA and the different AIA passbands. This is depicted in Fig.~\ref{fig10}, where the scatter plots between the entire FoV power maps of ALMA and other passbands, in the six period bins, are shown. The value of cross-correlation coefficients (cc) between the respective power maps are also mentioned in the scatter plots. The scatter plots show the absence of any kind of statistical trend or relation of ALMA oscillation power with that of the IRIS and AIA. The corresponding scatter plots separately for the bright plage region and the peripheral region also show very low values of the cross-correlation, less than 12\% for most of the cases. All the analyses above strongly indicate the lack of any association between oscillations observed with ALMA and that observed with IRIS and AIA, the possible reasons of which are discussed in the next section.

\section{Discussion}
\label{disc}

    We speculate that the low correlations are possibly due to the large variations in formation height of the millimeter continuum about 1.25\,mm observed by ALMA Band-6. Such large formation-height variations may also be present in other wide-band images analyzed here. As mentioned in Sect.~\ref{intro} (and citations therein), the numerical models have predicted ALMA to sample the mid-to-high chromosphere, although the exact heights of formation of the ALMA signal are still to be uncovered. \citet{Loukitcheva2015} employed Bifrost numerical simulations \citep{Gudiksen2011} to discuss the effects of the broad contribution functions of different ALMA bands in general. They have discussed that the millimeter radiation that results in the brightness temperatures deduced from ALMA observations, represent the integrated physical state of the atmosphere over an extended height range ($\sim$\,500\,km). They have shown that depending on the underlying magnetic structure, the contribution functions of different ALMA bands can have distinct multiple peaks, representing well separated different height ranges. Their analysis also depicts that in the regions with strong magnetic field, the millimeter observations tend to map comparatively higher atmospheric layers than the surrounding weaker magnetic field locations. Recently \citet{Eklund2021} have shown similar results, where they have also investigated the different ALMA bands in the Bifrost simulations and provided the distributions of their formation heights in various solar regions with different levels of chromospheric magnetic flux. They found a range of about 500--2000\,km for the synthetic ALMA Band-6 images, with a larger peak occurring towards the higher end in the more magnetically active regions. Such large variations in formation height can create strong temporal modulations that could destroy the oscillatory signals at particular periods (e.g., at around 3 min). 
    
    \citet{Rutten2017} and \citet{Jafarzadeh2021} discussed the role that magnetic-field strength and topology may play in the propagation of magnetoacoustic waves in the chromosphere. In particular, \citet{Jafarzadeh2021} performed a statistical study of ALMA oscillations using 10 different datasets (from both band 3 and 6) and found that the behavior of ALMA oscillations is only similar to that of AIA 1600\,\AA~in the most magnetically quiescent datasets, with the presence of 3--5\,min oscillations, whereas the datasets influenced by higher magnetic flux (i.e., presence of active and/or plage regions inside and/or in the immediate vicinity of the observed field of views) showed a lack of 3--5\,min oscillations and average power peaks at longer periods. In the present study, while we also find no clear 3--5\,min oscillations in the ALMA Band-6 observations, such oscillations were observed in both AIA 1600\,\AA~and IRIS SJI 2796\,\AA~datasets. While the former samples heights corresponding to the low chromosphere/temperature minimum, the latter is supposedly formed around the middle chromosphere. Thus, if the magnetic canopy would act as an obscuring effect, or if the large magnetic strength could suppress the power at shorter periods as discussed by \citet{Jafarzadeh2021} for the ALMA observations, those should also similarly influence other chromospheric observations, such as those with IRIS SJI 2796\,\AA, if they were both formed at similar heights. However, as noted above, the exact heights of formation of the ALMA observations are still unknown. In addition, the IRIS SJIs are obtained using relatively broadband filters and may include contributions from a large range of heights. How the heights of formation of the ALMA Band-6 and IRIS SJI 2796\,\AA~are compared could serve as the key to understand the discrepancy. Also, further studies using narrow-band observations sampling various heights through the solar chromosphere can clarify the speculations mentioned above.

    Recent studies have also indicated that ALMA observations may have some contributions from the transition region and corona  \citep{Wedemeyer2020, Chintzoglou2021A, Chintzoglou2021B}. It is thus likely that the ALMA observations have the simultaneous oscillatory contributions from a range of geometrical heights from the chromosphere to the corona. Furthermore, it is important to note that ALMA observations map the brightness temperature fluctuations, which is estimated from the continuum intensity under the Rayleigh-Jeans limit. This is in contrast to the intensity fluctuations observed by broadband filter images by IRIS and AIA, which depend on the collective behavior of the plasma temperature and density. The temperature and density oscillations may have complex phase relations depending on the frequency and wave mode, and the thermodynamical conditions in the different atmospheric layers, with adiabatic conditions to be dominant in the chromosphere and isothermal conditions in the corona \citep{Severino13, Vandoor11}. With the possibility of ALMA sampling these very different thermodynamical environments simultaneously, with non-equilibrium effects prevalent in the chromosphere and transition region, the coupling between ALMA temperature oscillations and IRIS/AIA intensity oscillations may become complex, which could explain the observed lack of association between ALMA temperature and IRIS/AIA intensity oscillations.

\section{Conclusions}
\label{conc}

    This article showcases a statistical comparison of the power distribution of oscillations in a plage region, spanning the solar atmosphere from the chromosphere to the lower corona,  utilizing coordinated observations from ALMA, IRIS, and SDO. We performed Lomb-Scargle transforms to statistically characterize the oscillations in ALMA Band-6 (1.25\,mm), IRIS 2796\,\AA\,SJI and AIA 1600\,\AA, 304\,\AA, 131\,\AA, 171\,\AA, 193\,\AA\,and 211\,\AA ~observations. The distribution of the dominant period reveals the presence of oscillations over a wide range of periods (up to 35\,min), with 12--14\,min to be prominent in ALMA and AIA coronal passbands (Figs.~\ref{fig3} and \ref{fig4} and \ref{fig5}). On the other hand, the chromospheric and transition region passbands of IRIS and AIA show a prominence of shorter periods of about 4--7\,min. It is worth noting that here we are discussing about the global dominant behavior of oscillations, whilst a large range of oscillation periods can be dominant in some of the individual pixels/features, as shown in the dominant period maps in Fig.~\ref{fig5}. The detection of high-frequency oscillations (<\,2\,min) have also been reported by \citet{Guevara2021} and \citet{Eklund2020} in some specific features in ALMA Band-3 observations.
    
   The ALMA oscillations in the plage region studied here are observed not to be associated with those observed by IRIS SJI and AIA. As shown in the scatter plots in Figs.~\ref{fig6},~\ref{fig7},~\ref{fig8} and \ref{fig10}, the ALMA dominant period map and power maps over the range of periods (up to 20\,min) do not show any spatial correlation with the respective ones from the other passbands, with all the cross-correlation coefficients to be less than 0.12. A better understanding about the range of formation height of the millimeter continuum observed by ALMA will help to explain the specific reasons for such non-associated behavior of ALMA oscillations with the IRIS and AIA oscillations. Additionally, the complex interrelation between the nature of ALMA temperature oscillations with the IRIS/AIA intensity oscillations need to be explored in detail. Further statistical investigations using multiple coordinated observations spanning a variety of solar features in quiet sun, coronal holes and active regions, complemented with numerical models, should help to better understand the nature of the oscillations observed with ALMA, and thus its association with other observations.

\begin{acknowledgements}
   This work is supported by the Research Council of Norway through its Centres of Excellence scheme, project number 262622 (Rosseland Centre for Solar Physics), and by the SolarALMA project, which has received funding from the European Research Council (ERC) under the European Union’s Horizon 2020 research and innovation programme (Grant agreement No. 682462). ALMA is a partnership of ESO (representing its member states), NSF (USA) and NINS (Japan), together with NRC (Canada), MOST and ASIAA (Taiwan), and KASI (Republic of Korea), in co-operation with the Republic of Chile. The Joint ALMA Observatory is operated by ESO, AUI/NRAO and NAOJ. IRIS is a NASA small explorer mission developed and operated by Lockheed Martin Solar and Astrophysics Laboratory (LMSAL) with mission operations executed at NASA Ames Research Center and major contributions to downlink communications funded by ESA and the Norwegian Space Centre. SDO is a mission for NASA's Living With a Star program, and data are provided by courtesy of NASA/SDO and the AIA and HMI science teams. Authors thank Luc Rouppe van der Voort for his inputs that helped to improve the manuscript. This research has made use of NASA’s Astrophysics Data System.    
\end{acknowledgements}

%
%

\bibliographystyle{aa} 
\bibliography{power_dist} 

\clearpage

\begin{appendix}

\section{Bright plage and peripheral region}

\label{apndxA}

\begin{figure}[htbp]
\centering
	\includegraphics[width=0.45\textwidth]{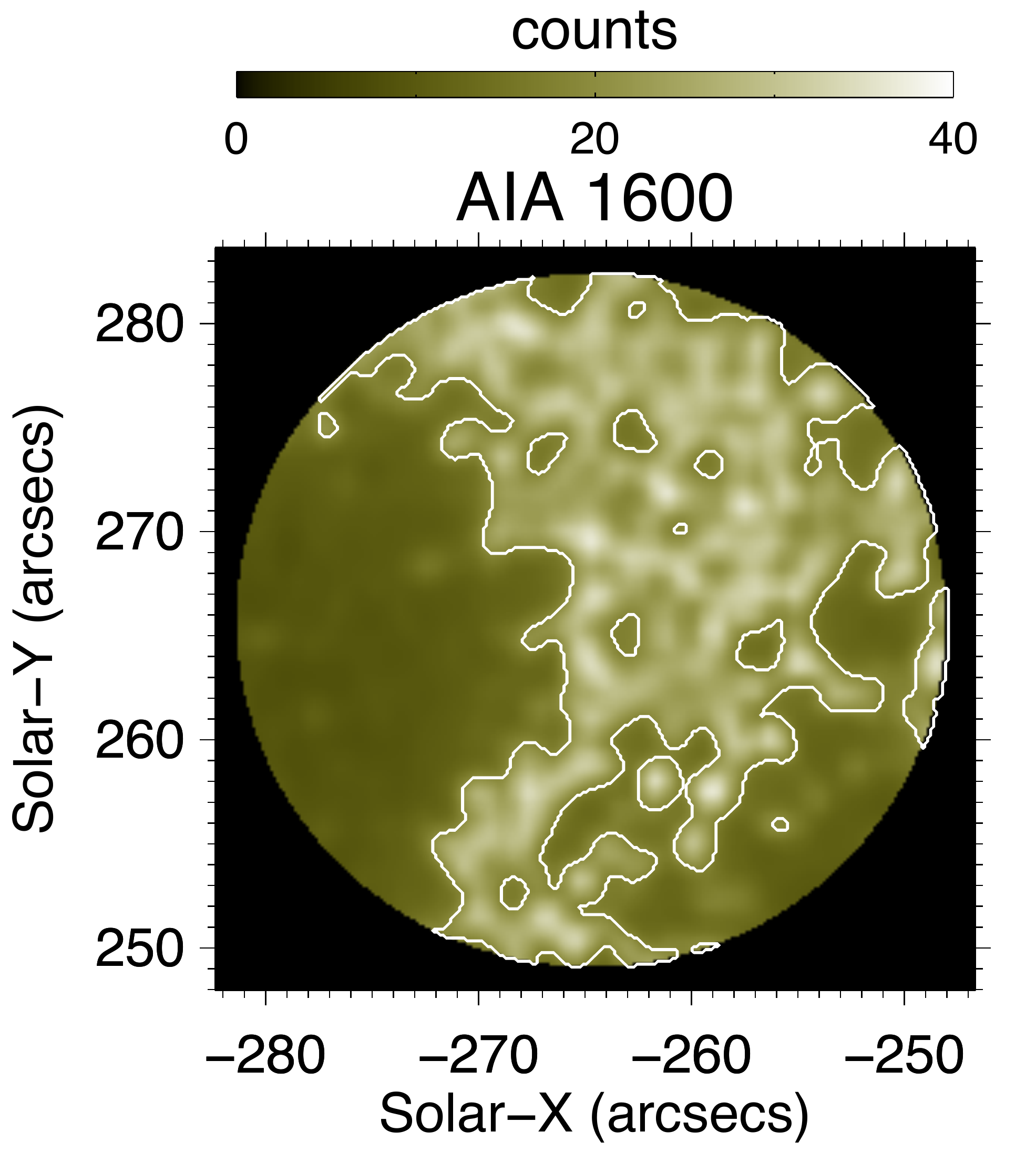}
	
	\caption{The time-averaged AIA 1600\,\AA~intensity image of the studied FoV. Overplotted 
	are the \textit{white} contours that enclose the bright plage region. The surrounding area (outside the contours) is the peripheral region.}
	
	\label{figA1}
\end{figure}

\begin{figure*}[htbp]
	\centering
	
	\includegraphics[width=0.90\textwidth]{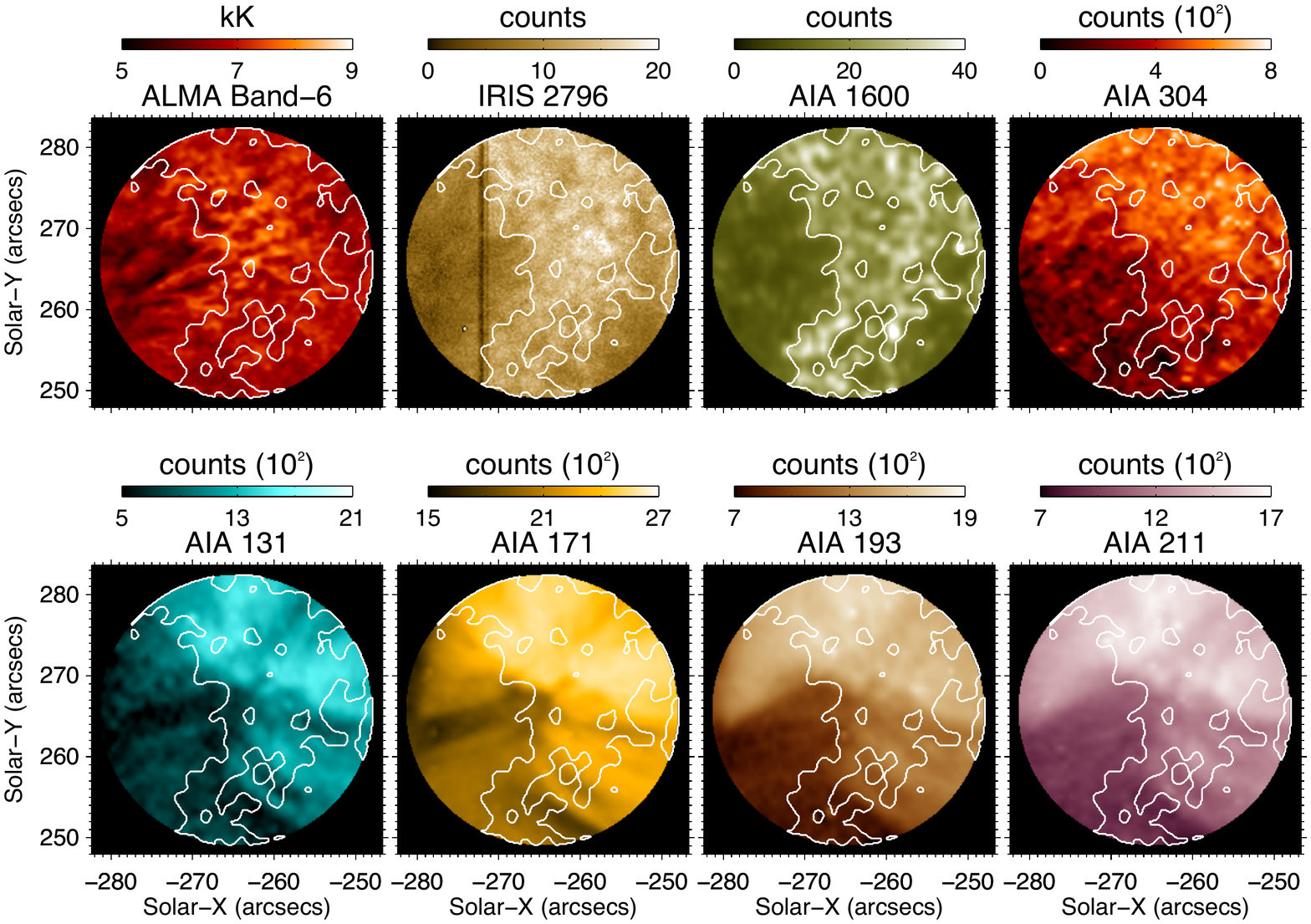}
	
	\caption{Representative images of the studied FoV (from ALMA Band-6, IRIS SJI\,2796\,\AA~, and different AIA channels as indicated on top of the panels) at the start time of the observations. Overplotted are the \textit{white} contours which separate the bright plage region from the peripheral region.}
	
	\label{figA2}
\end{figure*}

As mentioned in scetion~\ref{sc}, we partitioned the FoV into two sub-regions, the bright plage region and the peripheral region, based on the intensity threshold from AIA 1600\,\AA~observations. From the time-averaged AIA 1600\,\AA~intensity image, we select the locations with more than average counts (average over the FoV) as the bright plage region. The remaining region of the FoV is assigned to belong to the peripheral region. Figure~\ref{figA1} shows the time-averaged AIA 1600\,\AA~image, where the \textit{white} contours enclose the bright plage region, and the surrounding area (outside the contours) is the peripheral region. One representative image from each of the passband studied is shown in Fig.~\ref{figA2} (same as Fig.~\ref{fig1}), with the \textit{white} contours separating the bright plage region from the peripheral region.

\end{appendix}

\end{document}